\documentclass[%
    article,
    twocolumn,
preprintnumbers,
bibnotes,amsmath,amssymb,
aps,
pre,
floatfix,
]{revtex4-2}

\usepackage{graphicx}
\usepackage{dcolumn}
\usepackage{pifont}
\usepackage{tikz}
\usepackage{circuitikz}
\usepackage{mathtools}
\usepackage{bm}
\usepackage{amsmath,amssymb}
\usepackage{hyperref}
\usepackage{mathtools}
\usepackage{lipsum}
\usepackage{booktabs}
\usepackage{subfigure}
\usepackage{stmaryrd}

\makeatletter
\pgfcircdeclarebipolescaled{instruments}
{
    \anchor{text}{\pgfextracty{\pgf@circ@res@up}{\northeast}
        \pgfpoint{-.5\wd\pgfnodeparttextbox}{
            \dimexpr.5\dp\pgfnodeparttextbox+.5\ht\pgfnodeparttextbox+\pgf@circ@res@up\relax
        }
    }
}
{\ctikzvalof{bipoles/oscope/height}}
{jj}
{\ctikzvalof{bipoles/oscope/height}}
{\ctikzvalof{bipoles/oscope/width}}
{
    \pgf@circ@setlinewidth{bipoles}{\pgfstartlinewidth}
    \pgfextracty{\pgf@circ@res@up}{\northeast}
    \pgfextractx{\pgf@circ@res@right}{\northeast}
    \pgfextractx{\pgf@circ@res@left}{\southwest}
    \pgfextracty{\pgf@circ@res@down}{\southwest}
    \pgfmathsetlength{\pgf@circ@res@step}{0.25*\pgf@circ@res@up}
    \pgfscope
        \pgfpathrectanglecorners{\pgfpoint{\pgf@circ@res@left}{\pgf@circ@res@down}}{\pgfpoint{\pgf@circ@res@right}{\pgf@circ@res@up}}
        \pgf@circ@draworfill
    \endpgfscope
    \pgfscope
      \pgfpathmoveto{\pgfpoint{\pgf@circ@res@left}{\pgf@circ@res@up}}%
      \pgfpathlineto{\pgfpoint{\pgf@circ@res@right}{\pgf@circ@res@down}}%
      \pgfpathmoveto{\pgfpoint{\pgf@circ@res@right}{\pgf@circ@res@up}}%
      \pgfpathlineto{\pgfpoint{\pgf@circ@res@left}{\pgf@circ@res@down}}%
      \pgfusepath{draw}
    \endpgfscope
}
\def\pgf@circ@jj@path#1{\pgf@circ@bipole@path{jj}{#1}}
\tikzset{jj/.style = {\circuitikzbasekey, /tikz/to path=\pgf@circ@jj@path, l=#1}}

\newcommand{\mrm}[1]{\mathrm{#1}}

\newcommand{\dUOneOne}{\Delta U_{11}'}
\newcommand{\dUZeroOne}{\Delta U_{01}'}
\newcommand{\dUOneZero}{\Delta U_{10}'}
\newcommand{\dUZeroZero}{\Delta U_{00}'}

\newcommand{\initPos}{\varphi_2^o}

\newcommand{\tCE}{t_{\mathrm{CE}}}
\newcommand{\tDwell}{t_{\mathrm{d}}}
\newcommand{\EscapeRateIndexOnly}{l}
\newcommand{\GammaIndexed}{\Gamma_l}
\newcommand{\PlasmaFreqIndexed}{\omega_{p,l}}
\newcommand{\tDwellIndexed}{t_{\mathrm{d},l}}
\newcommand{\tildeOmega}{\widetilde{\omega}}

\newcommand{\vpxdcIndexI}{\varphi_{i\mathrm{xdc}}}
\newcommand{\vpdcIndexI}{\varphi_{i\mathrm{dc}}}

\newcommand{\vpxdcIndexOnly}{i}

\newcommand{\vpxdcTwo}{\varphi_{2\mathrm{xdc}}}

\newcommand{\vpxOne}{\varphi_{1\mathrm{x}}}
\newcommand{\vpxTwo}{\varphi_{2\mathrm{x}}}

\newcommand{\vpxBarrierOffset}{\varphi_{j\mathrm{x}}}

\newcommand{\vpxBiasOffset}{\varphi_{i\mathrm{x}}}

\newcommand{\mInitX}{m_{x}(t=t_i)}
\newcommand{\mInitY}{m_{y}(t=t_i)}
\newcommand{\mFinX}{m_{x}(t=t_f)}
\newcommand{\mFinY}{m_{y}(t=t_f)}

\newcommand{\tX}{\mathtt{X}}
\newcommand{\tY}{\mathtt{Y}}
\newcommand{\tXprime}{\mathtt{X}^\prime}
\newcommand{\tYprime}{\mathtt{Y}^\prime}

\newcommand{\CEComp}{(I_{\tX}, C_{\overline{\tX}} E_1)}

\newcommand{\IdNotation}{I_\rho}
\newcommand{\IdNotationSymbol}{\rho}

\newcommand{\IdOpX}{I_{\tX}}

\newcommand{\ContNotation}{C_\sigma}
\newcommand{\ContNotationSymbol}{\sigma}

\newcommand{\ErasNotation}{E_A}

\newcommand{\pOneMin}{\phi_1^{*}}
\newcommand{\pTwoMin}{\phi_2^{*}}

\newcommand{\figNumeric}[2]{#1(#2)}
\newcommand{\figNumericNoSpace}[2]{#1(#2)}

\newcommand{\CEOne}{(I_{\tX}, C_{\overline{\tX}} E_1)}
\newcommand{\CETwo}{(I_\tX, C_{\tX} E_1)}
\newcommand{\CEThree}{(C_\tY E_1, I_\tY)}
\newcommand{\CEFour}{(C_{\overline{\tY}} E_1, I_\tY)}
\newcommand{\CEFive}{(I_\tX, C_{\overline{\tX}} E_0)}
\newcommand{\CESix}{(I_\tX, C_\tX E_0)}
\newcommand{\CESeven}{(C_\tY E_0, I_\tY)}
\newcommand{\CEEight}{(C_{\overline{\tY}} E_0, I_\tY)}

\newcommand{\PartialCompNotation}{P_{\rho^\prime}[(\tX, \tY)]}
\newcommand{\CompleteCompNotation}{C[(\tX, \tY)]}

\newcommand{\PartialCompExample}[1]{P_{#1}[(\tX, \tY)]}

\newcommand{\ModIversonBracket}[1]{\llbracket #1 \rrbracket}
\newcommand{\ModIversonBracketNotation}{\llbracket P \rrbracket}

\newcommand{\xyplane}{$x$\nobreakdash--$y$~plane}
\newcommand{\vphiOnevphiTwoPlane}{$\varphi_1$\nobreakdash--$\varphi_2$~plane}

\newcommand{\arxiv}[1]{\href{http://arxiv.org/abs/#1}{\texttt{arXiv}:#1}}
\newcommand{\CE}{\textit{control erase} }

\begin{document}

\preprint{\arxiv{2406.12153}}

\title{\texorpdfstring{Controlled Erasure as a Building Block for \\
Universal Thermodynamically-Robust Superconducting Computing}{Coupled SQUIDS as Building Blocks for Universal Computing}}

\author{Christian Z. Pratt}%
 \email{czpratt@ucdavis.edu}

\author{Kyle J. Ray}
    \email{kjray@ucdavis.edu}
 
\author{James P. Crutchfield}
    \email{chaos@ucdavis.edu}
\affiliation{%
Complexity Sciences Center and Department of Physics and Astronomy, University of California, Davis, One Shields Avenue, Davis, CA 95616
}%

\date{\today}

\begin{abstract}

Reducing the energy inefficiency of conventional CMOS-based computing devices---which rely on logically irreversible gates to process information---remains both a fundamental engineering challenge and a practical social challenge of increasing importance. We extend an alternative computing paradigm that manipulates microstate distributions to store information in the metastable minima determined by an effective potential energy landscape. These minima serve as mesoscopic memories that are manipulated by a dynamic landscape to perform information processing. Central to our results is the \textit{control erase} (CE) protocol that controls the landscape's metastable minima to determine whether information is preserved or erased. Importantly, successive protocol executions can implement a NAND gate---a logically-irreversible universal logic gate. We show how to practically implement this in a device created by two inductively-coupled superconducting quantum interference devices (SQUIDs). We identify circuit parameter ranges that give rise to effective CEs and establish the device's robustness against logical errors. These SQUID-based logical devices are capable of operating above GHz frequencies and at the $k_B T$ energy scale. Due to this, optimized devices and associated protocols provide a universal-computation substrate that is both computationally fast and energy efficient.

\end{abstract}
\maketitle

\section{Introduction}\label{sec:Introduction}

Conventional classical computing systems harness irreversible logic gates---e.g., NAND gates---to process information. Irreversibility arises from erasing input information to create desired output states, coming at a minimum theoretical heat dissipation cost of $k_B T \ln 2$ \cite{Landauer_1961} per erasure. During their operation, current conventional CMOS-based computational devices generate $\mathcal{O}(10^4)$ times more heat than this minimum cost \cite{Freitas_Delvenne_Esposito_2021, Gao_Limmer_2021, Takeuchi_2022, Intel}. On the social scale, energy consumption for computational purposes is projected to reach 20\% of global energy demand by 2030 \cite{Jones_2018}. In light of this, it is time to investigate alternative strategies and substrates that give rise to markedly-more energy-efficient universal computation.

One strategy implied in Landauer's seminal 1961 result \cite{Landauer_1961, Bennett_1982, Sekimoto_1997, Plenio_Vitelli_2001, Blickle_Speck_Helden_Seifert_Bechinger_2006, Jun_Gavrilov_Bechhoefer_2014, Riechers_2019, Riechers_Boyd_Wimsatt_Crutchfield_2020, Boyd_Patra_Jarzynski_Crutchfield_2022} focuses on manipulating metastable energy minima in an effective potential energy landscape. Coarse-graining the microstate phase-space surrounding the minima yields long-lived mesoscopic memory states that correspond to the logical $0$s and $1$s for binary computation. From this perspective, a computation is implemented by controlling memory-state dynamics through changes in the landscape. Altogether, a computation is then a map between initial and final memory states. This ``energy first'' perspective of computation allows for careful analysis and prediction of a given logic gate's performance and efficiency \cite{Landauer_1961, Bennett_1982, Jun_Gavrilov_Bechhoefer_2014, Boyd_Patra_Jarzynski_Crutchfield_2022}.

Elaborating on this framework, Sec. \ref{sec:PropertiesPotentialEnergyLandscape} first describes a family of potential energy landscapes capable of storing two bits of information. Section \ref{sec:ControlErasureProtocolSetup} introduces a family of \CE (CE) protocols---two-dimensional generalizations of single-bit erasure protocols. Section \ref{sec:DeviceImplementation} then presents a device constructed from two inductively coupled superconducting quantum interference devices (SQUIDs) \cite{Harada_Goto_Miyamoto_1987, Han_Lapointe_Lukens_1989, Han_1992_rf, Han_Lapointe_Lukens_1992, Harris_Johnson_Han_2008, Pratt_Ray_Crutchfield_2024}. We demonstrate that this device implements the required family of potential energy landscapes and executes CE protocols by changing its circuit parameters. Sections \ref{sec:FindingEffectiveCEProtocols}-\ref{sec:CERobustness} explore the effectiveness of the SQUID CE implementation as a function of circuit parameters. Notably, serial executions of CEs produce a robust NAND gate, as detailed in Sec. \ref{sec:NANDGateApplication}. Executing robust NAND gates with this device supports extremely low-energy universal computation.

\section{Physical Computing via Metastable Landscapes}
\label{sec:PropertiesPotentialEnergyLandscape}

Consider a potential energy landscape that exhibits several energy minima supported via an underlying physical substrate which itself is connected to a thermal environment that introduces damping and noise into its dynamics. The minima are separated by energy barriers, whose heights are substantially larger than the thermal energy $k_B T$. The thermal environment quickly induces distributions of microstates to settle into local equilibria in the phase-space regions surrounding these wells. Noise perturbs the microstates within their respective regions; however, it is unlikely to drive the microstates between these regions on a timescale that scales exponentialy with the energy barrier height. The energy barriers prevent mixing between these regions except on these very long timescales. As a result, the minima serve to support long-lived mesoscopic system states---metastable memory states. Manipulating them with the potential's dynamics corresponds to information processing. 

The following details physically-embedded computations via landscape control in this setting \cite{Landauer_1961, Reguera_2005, Esposito_2012, Seifert_2018, Riechers_Boyd_Wimsatt_Crutchfield_2020, Ray_Boyd_Wimsatt_Crutchfield_2021}. Specifically, these computations are stochastic mappings from the set of initial to final memory states $\mathcal{M}$. Here, we introduce the \CE (CE) protocol over two input bits. One way of embedding 2-bit computations is to control a two-dimensional potential with a minimum in each quadrant. The memory states are distinguished using the $x$ and $y$ axes:
\begin{align}
\label{eq:mem state instantiation rule}
    \text{first [second] bit} = \begin{cases}
        & 0 \text{\; if \; } x \ [y] < 0 ~, \\
        & 1 \text{\; if \;  }x \ [y] > 0 ~.
    \end{cases} 
\end{align}

\begin{figure}[t]
    \centering
    \includegraphics[width=\linewidth]{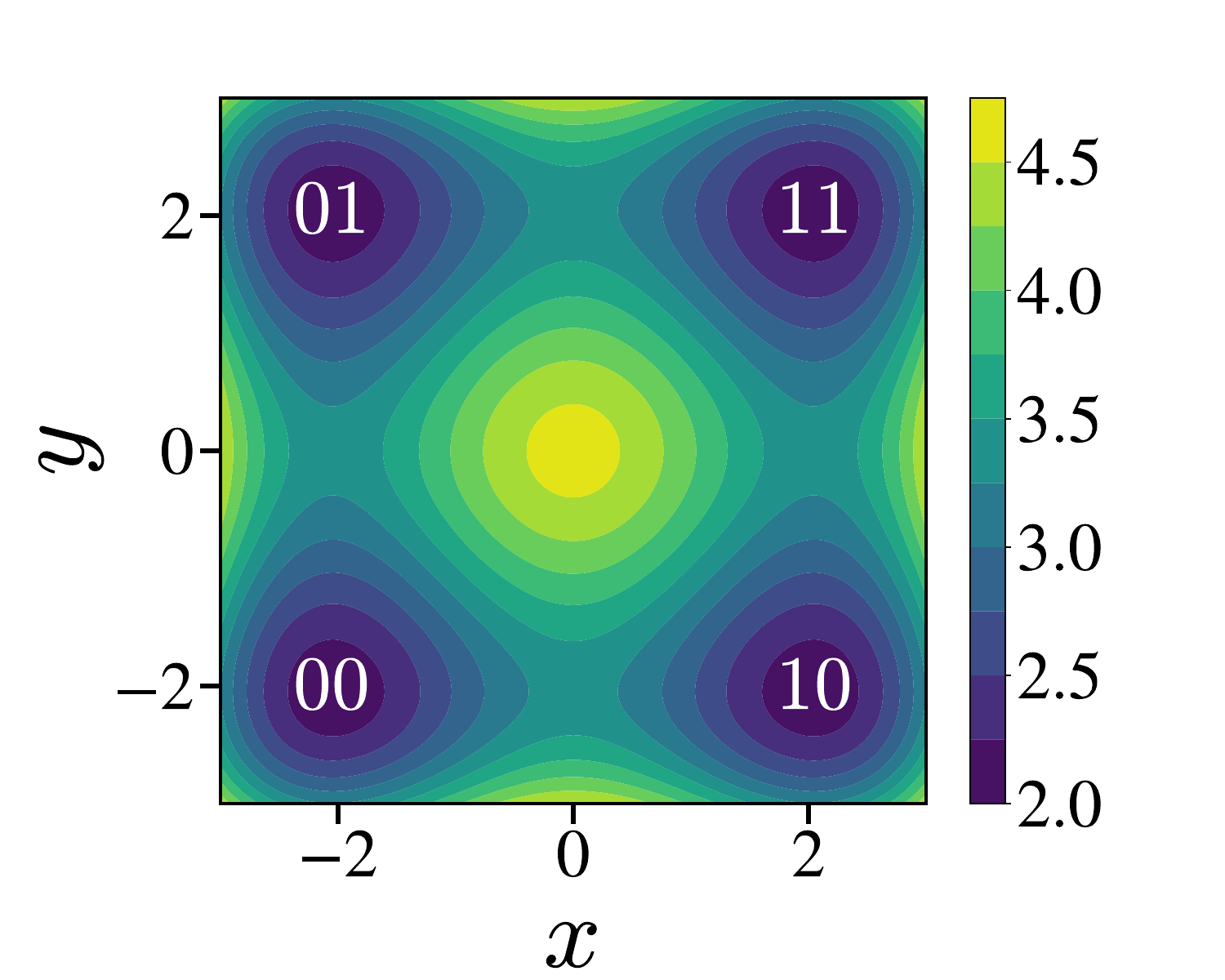}
    \caption{Example 2-bit memory mesoscopic-state instantiation in a four well potential. Regions surrounding energy minima provide metastable information storage, whose bit value is assigned according to Eq. (\ref{eq:mem state instantiation rule}). Controlling this energy landscape manipulates the memory state's dynamics, thereby implementing information processing.}
    \label{fig:landscape mem state instantiation}
\end{figure}

Figure \ref{fig:landscape mem state instantiation} illustrates a quadruple-well potential with an example memory-state instantiation chosen by the locations of the minima according to Eq. (\ref{eq:mem state instantiation rule}). The computational operations to be performed on the potential---i.e., the deterministic transformation of the potential that maps information from initial to final metastable memory states---is referred to as a \emph{physical protocol}. 

Last, we formally define metastable memory states in the \xyplane\ and how they evolve under a computational protocol. Over the time interval $t \in [t_i, t_f]$ and given an initial two-bit metastable memory state $(\tX, \tY) \coloneqq (\mInitX, \mInitY) \in \mathcal{M}$, the final memory state $(\tXprime, \tYprime) \coloneqq (\mFinX, \mFinY)$ is determined by the conditional probability $\mathbf{p} = \Pr((\tXprime, \tYprime) | (\tX, \tY))$. With this, the final microstate distribution $\Vec{p}(t=t_f)$ is updated through $\Vec{p}(t=t_f) = \mathbf{p} \, \Vec{p}(t=t_i)$. A deterministic logic gate is successfully performed when these conditional probabilities are very close to either $0$ or $1$. 

\section{Control Erase Protocol}\label{sec:ControlErasureProtocolSetup}

The NAND gate is a binary logical gate that takes in two input bits, meaning there are four possible input states---$00$, $01$, $10$, and $11$---and one output bit---either $0$ or $1$. Specifically, the output state is $1$ for all input states except if both input bits are $1$, in which case the output is $0$. 

Consider reading the output state to be $1$ with no knowledge of a NAND gate's input state. Since there are three possible inputs that can produce this single output state, inferring which specific two-bit input state led to the output bit $1$ is impossible. The logical impossibility manifests as a physically-irreversible erasure of information associated with phase-space contraction at the microscopic scale. In contrast, if the output was $0$, then the input is trivially known to be $11$ and no phase-space contraction is entailed.

The above illustrates the NAND's underlying information processing task: Preserve some input information, while erasing the rest. In this way, performing a NAND gate requires the ability to carry out controlled erasure protocols over the space of memory states. 

\begin{figure}[t]
    \centering
    \includegraphics[width=0.95\linewidth]{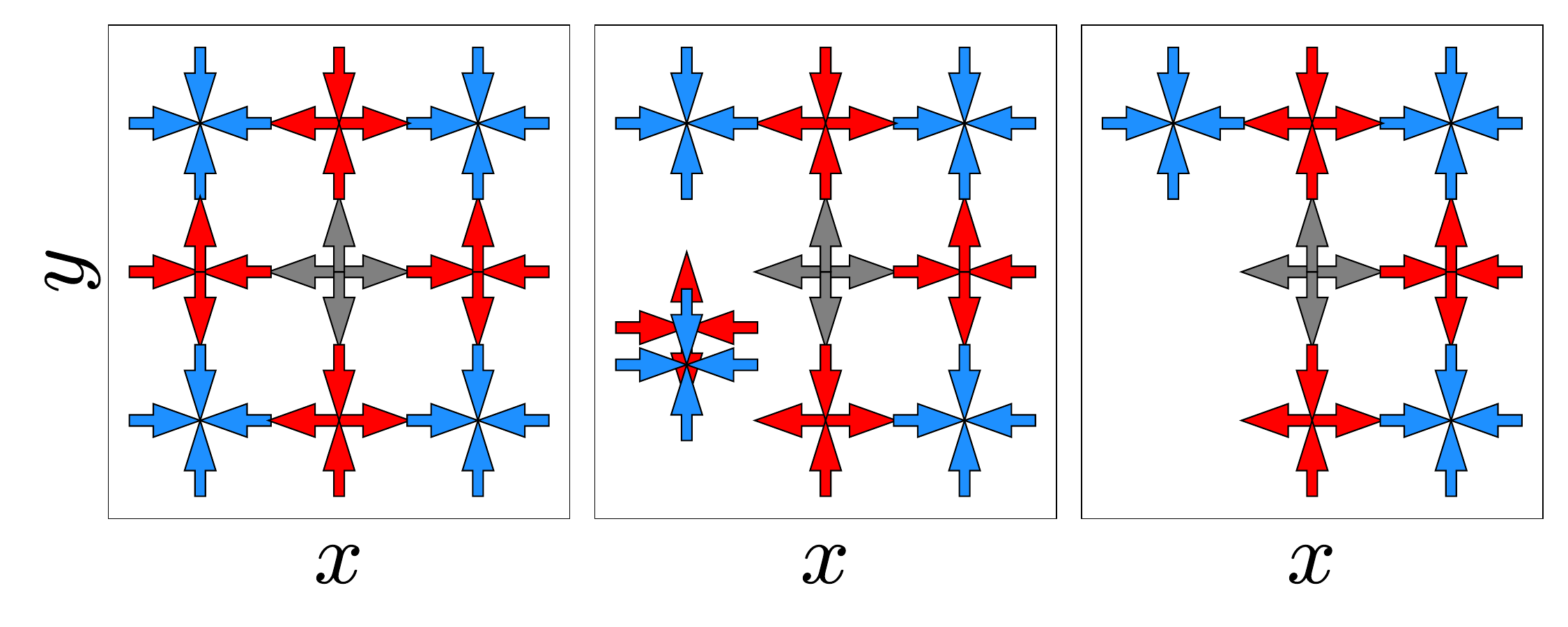}
    \caption{Control erase (CE) protocol illustrated via the \emph{dynamical skeleton} of the potential energy landscape in Fig. \ref{fig:landscape mem state instantiation}, showing only the potential's fixed points and corresponding local flow fields. Stable minima (unstable saddle) fixed points are colored blue (red), while grey indicates a local maximum. This CE is executed with respect to the $y$ axis: The states contained on the negative $x$ half plane undergo an erasure of $\tY$ to the $1$ state, while those on the positive $x$ half plane do not change. This results in the third quadrant's stored information being erased into the second quadrant. (Left) Initial potential energy landscape configuration. (Center) The stable and unstable fixed points in quadrant III approach each other just before annihilation. (Right) Immediately after fixed point annihilation in the third quadrant. If this landscape is held long enough, all information initially stored in quadrant III will be erased to quadrant II. Importantly, all other fixed points are maintained during this time. One completes the CE protocol cycle by bringing the potential back to its starting configuration.}
    \label{fig:erasure skeletons}
\end{figure}

Erasure protocols in one-dimensional systems have been studied previously \cite{Landauer_1961, Dillenschneider_Lutz_2009, Jun_Gavrilov_Bechhoefer_2014, Talukdar_Bhaban_Salapaka_2017, Riechers_Boyd_Wimsatt_Crutchfield_2020, Saira_2020, Proesmans_Ehrich_Bechhoefer_2020, Wimsatt_Saira_Boyd_Matheny_Han_Roukes_Crutchfield_2021}. The following extends this to an erasure protocol via the two-dimensional potential shown in Fig. \ref{fig:landscape mem state instantiation}. The additional freedom given by the second dimension permits control over what information is preserved and erased. Via that control, a NAND operation can be performed. To accomplish this, we introduce the \CE (CE) protocol. Figure \ref{fig:erasure skeletons} illustrates a CE protocol via the potential's \emph{skeleton}, viewing the potential from a dynamical systems perspective---i.e., via its fixed points and local flow fields in the microscopic state space. In this example, the negative $x$ half-plane executes an erasure protocol, while the positive $x$ half-plane remains unchanged.

\begin{figure*}
        \centering
        \includegraphics[width=\linewidth]{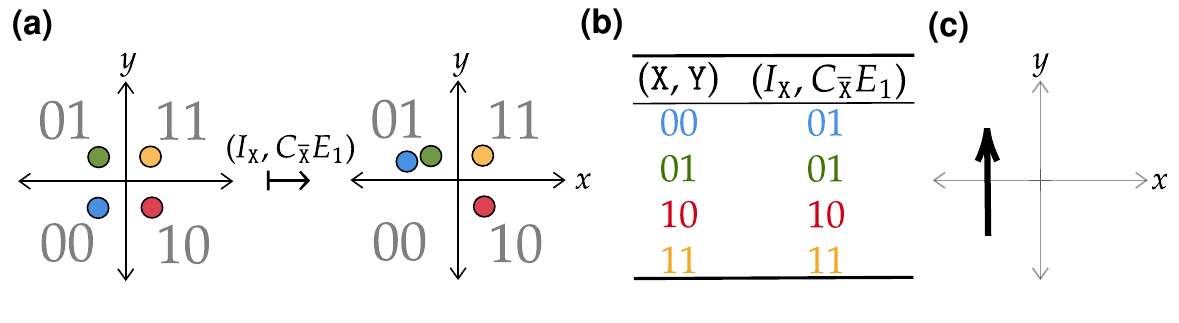}
        \caption{(a) State mapping executed by the control erasure $(\tXprime, \tYprime) = \CEComp$. Each quadrant contains a coarse-grained metastable memory state, defined by Eq. (\ref{eq:mem state instantiation rule}). Every circle represents a distribution of microstates, i.e. information, and is colored based on its original memory-state instantiation. (b) CE truth table. The overline above state $\tX$, in the symbol $C_{\overline{\tX}}$, indicates controlling on the negation of $\tX$, erasing $\tY$ to $1$ only when $\tX=0$. (c) Arrow-based notation illustrating the $\CEComp$ CE protocol. This depiction is also used in Sec. \ref{sec:NANDGateApplication}. Fig. \ref{fig:AllControlErasureProtocols} shows all CEs of interest.}
    \label{fig:CE example tableau}
\end{figure*}

Executing a given CE protocol requires three binary choices, illustrated via examples:
\begin{enumerate}
    \item First, choose which of the two input bits to operate on. For example, erasing information along the $y$ axis requires operating only on the second bit so that $\tYprime\neq\tY$. This implies that the first input bit is preserved, i.e., $\tXprime=\tX$. However, this does not fully specify the pair of states that involved in the CE: $00$ and $01$, or $10$ and $11$, are both candidate pairs for a $\tYprime\neq\tY$ operation.
    \item Next, specify which pair of two-bit memory states are involved in the erasure. Suppose we select the states $10$ and $11$: Doing this \textit{controls} the erasure operation on $\tX$ such that information is erased in $\tY$ only if $\tX = 1$. Otherwise, if $\tX = 0$, the information stored in $\tY$ is preserved. We denote this choice as a function $CE(\tX, \tY)$ that is an identity on $\tY$ if $\tX=0$ and an erasure of $\tY$ if $\tX=1$.
    \item However, we also need to choose whether $\tY$ is erased to the $0$ or the $1$ state. For example, if we erase to the state $1$, we augment our notation to be $C E_1(\tX,\tY)$.
\end{enumerate}

These three binary choices produce the eight CE protocols of interest shown in Fig. \ref{fig:AllControlErasureProtocols}. All said, the choices above give us $(\tXprime, \tYprime) = (I(\tX), C E_1(\tX, \tY))$, where $I(\tX)$ indicates an identity operation on the $\tX$ input bit. For legibility, we introduce a condensed notation that suppresses the functional dependencies: $(\tXprime, \tYprime) = (\IdOpX , C_{\tX} E_1)$. The following relies on this notation.

Figure \ref{fig:CE example tableau} shows multiple representations of another CE: $\CEComp$. This also erases information stored in $\tY$, but instead of erasing the information if $\tX=1$, the erasure happens if $\tX=0$. In other words, our control is $\overline{\tX}$ (NOT $\tX$), rather than $\tX$. Of particular importance is the arrow-based notation in Fig. \ref{fig:CE example tableau}, which is employed when detailing the NAND gate in Sec. \ref{sec:NANDGateApplication}. 

Finally, to generalize the CE notation, first we know that one of the two output states $\rho^\prime \in \{ \tXprime, \tYprime\}$ is created by the identity operation $\rho' = \IdNotation$, where $\rho\in \{ \tX, \tY \}$. The other output state is created by $\ContNotation \ErasNotation$. Here, $\ContNotationSymbol \in \{ \IdNotationSymbol, \ \overline{\IdNotationSymbol} \}$ indicates the control state for an erasure to the $A \in \{ 0,1 \}$ state. Note that the overline $\overline{\rho}$ above $\rho$ indicates a negation of the $\rho$ state. The eight CEs of interest are shown in Fig. \ref{fig:AllControlErasureProtocols} in App. \ref{app:CEs}.

\section{Device Implementation}
\label{sec:DeviceImplementation}

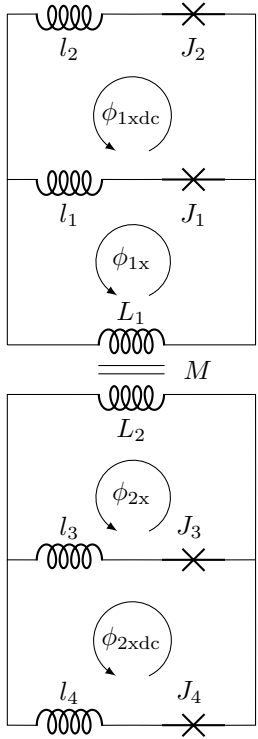
\begin{figure}[t]
    \begin{center}
     \begin{circuitikz}[scale=1.1]
    \tikzstyle{every node}=[font=\normalsize]

    \draw (0,0)
    to[L=$L_1$, label distance=0.0cm] (3,0)
    to[short] (3,2)
    to[barrier=$J_1$, label distance=-5cm] (1.5,2)
    to[L=$l_1$, label distance=0.0cm] (0,2)
    to[short] (0,0);

    \draw (3,2)
    to[short] (3,4)
    to[barrier=$J_2$, label distance=-5cm] (1.5,4)
    to[L=$l_2$] (0,4)
    to[short] (0,2);

    \draw (3,-0.6)
    to[L=$L_2$] (0,-0.6)
    to[short] (0,-2.6)
    to[L=$l_3$] (1.5, -2.6)
    
    to[barrier] (3,-2.6) 
    to[short] (3,-0.6);

    \draw (0,-2.6)
    to[short] (0,-4.6)
    to[L=$l_4$] (1.5,-4.6)
    to[barrier] (3,-4.6) 
    to[short] (3,-2.6);
    
      
    \draw[thin, ->, >=latex] (1.5,1.05)node{$\mathrm{\phi_{1\mathrm{x}}}$}  ++(-65:0.50) arc (-65:245:0.45);
      
    \draw[thin, ->, >=latex] (1.5,2.8)node{$\mathrm{\phi_{1\mathrm{xdc}}}$}  ++(-65:0.50) arc (-65:245:0.47);

    \draw[thin, ->, >=latex] (1.5,-1.8)node{$\mathrm{\phi_{2\mathrm{x}}}$}  ++(-65:0.50) arc (-65:245:0.45);

    \draw[thin, ->, >=latex] (1.5,-3.55)node{$\mathrm{\phi_{2\mathrm{xdc}}}$}  ++(-65:0.50) arc (-65:245:0.47);

    \draw[-] (1.1,-0.25) -- (1.9,-0.25) node{};
    \draw[-] (1.1,-0.35) -- (1.9,-0.35) node{};

    \node at (2.3,-0.3) {$M$};

    \node at (2.2, -2.2) {$J_3$};

    \node at (2.2, -4.2) {$J_4$};

    \end{circuitikz}
\caption{Two inductively coupled SQUIDs interacting via mutual inductance coupling constant $M$.}
\label{fig:universal computing circuit}
\end{center}
\end{figure}

Under the current computing paradigm, superconducting quantum interference devices (SQUIDs) are employed only as a physical platform for performing CE protocols. (To emphasize, other than zero-resistance supercurrents and the Josephson nonlinearity, no quantum-mechanical phenomena are used for information processing. And, low-temperate operation is relied on only insofar as it freezes out irrelevant thermal degrees of freedom.) We first briefly review related superconducting device physics. Then, we specify the device of interest---two inductively coupled SQUIDs, shown in Fig. \ref{fig:universal computing circuit}---whose potential energy landscape can dynamically execute CE protocols.

Historically, the \emph{quantum flux parametron} (QFP) was invented by Goto in 1985 \cite{Loe_Goto_1985, Harada_Goto_Miyamoto_1987} for classical computing applications on an energy-efficient superconducting platform. With the incentive of optimizing circuit parameters and operating on slower computational timescales to aid energy efficiency, Takeuchi et al introduced the \emph{adiabatic QFP} \cite{Takeuchi_Ozawa_Yamanashi_Yoshikawa_2013, Takeuchi_2022}. In contrast to our focus here, their circuit parameters were not optimized for thermodynamic performance, but were instead were determined via bifurcation theory. With a similar design construction as the QFP, but having the same equations of motion as the QFP, in 1989 Han et al introduced the \emph{variable $\beta$ radio frequency SQUID} \cite{Han_Lapointe_Lukens_1989, Han_Lapointe_Lukens_1992, Han_1992_rf}. This device was later named the \emph{compound Josephson junction radio frequency SQUID} (CJJ rf SQUID) \cite{Harris_Berkley_Johnson_2007, Harris_Johnson_Han_2008, Harris_Lanting_2009}. Its principle use was for investigating macroscopic quantum phenomena (MQP), not classical information processing. As such, our present goal---demonstrating universal classical information and storage---is not shared with the CJJ rf SQUID literature. 

This said, the circuit parameter values used in these earlier devices happen to be useful in Sec. \ref{sec:CERobustness} for evaluating the performance of our device of interest when performing CE protocols. Since QFPs and CJJ rf SQUIDs are both SQUIDs, we refer to the device in Fig. \ref{fig:universal computing circuit} as two inductively coupled SQUIDs. Reference \cite{Pratt_Ray_Crutchfield_2024} introduced and detailed the device in Fig. \ref{fig:universal computing circuit}. Importantly, it supports a potential equivalent to that of Fig. \ref{fig:landscape mem state instantiation}. To understand how it executes computations, the following gives a synopsis of the device's potential energy surface. Physically-grounded approximations and assumptions for executing a CE protocol are then detailed.

Figure \ref{fig:universal computing circuit} illustrates a device consisting of two inductively coupled SQUIDs. In this circuit, $L_i$ [$l_j$] indicate the radio frequency [direct current] SQUID inductances, while $J_j$ indicate Josephson junctions, all for which $i=1,2$, and $j= 1,2,3,4$. This device generates the following potential energy surface:
\begin{align}
U' = U/U_0 & = \dfrac{1}{2} (\varphi_1 -
\varphi_{1\mathrm{x}})^2 + \dfrac{1}{2} (\varphi_2 -
\varphi_{2\mathrm{x}})^2  \nonumber \\
&  \quad + \dfrac{\gamma_1}{2} (\varphi_{1\mathrm{dc}}
- \varphi_{1\mathrm{xdc}})^2 + \dfrac{\gamma_2}{2} (\varphi_{2\mathrm{dc}}
- \varphi_{2\mathrm{xdc}})^2  \nonumber \\
&  \quad - \beta_1 \cos \dfrac{\varphi_{1\mathrm{dc}}}{2} \cos
\varphi_1 - \beta_2 \cos \dfrac{\varphi_{2\mathrm{dc}}}{2} \cos
\varphi_2  \nonumber \\
&  \quad + \delta \beta_1 \sin \dfrac{\varphi_{1\mathrm{dc}}}{2} \sin \varphi_1  + \delta \beta_2 \sin \dfrac{\varphi_{2\mathrm{dc}}}{2} \sin \varphi_2  \nonumber \\
& \quad  + \mu(\varphi_1 - \varphi_{1\mrm{x}})(\varphi_2 - \varphi_{2\mrm{x}}) ~.
\label{eq:two bit potential without approximations}
\end{align}
Here, $\varphi_r = (2\pi \phi_r)/\Phi_0$ is the reduced flux variable of the $r$th flux variable,  while $\Phi_0$ is the flux quantum. We subsequently take $L \coloneqq L_1 = L_2$ and $l_{1(2)} \coloneqq l_{1(3)} = l_{2(4)}$. Next, $L_\alpha = \alpha L$ for which $\alpha = 1 - \mu^2$ such that $\mu = M/L$. $M$ is a tunable mutual inductance constant: While its experimental implementation---a SQUID coupler between the two SQUIDs---are discussed in Refs. \cite{Brink_Berkley_Yalowsky_2005, Harris_Berkley_Johnson_2007, Harris_Lanting_2009}, the coupler's dynamics will not be addressed here. With this, $U_0 = (\Phi_0/2\pi)^2/L_\alpha$ and $\gamma_{1(2)} = L_\alpha/2l_{1(2)}$. Last, $\beta_{1(2)} = 2\pi L_\alpha (I_{\mrm{c}2(4)} + I_{\mrm{c}1(3)})/\Phi_0$, and $\delta \beta_{1(2)} = 2\pi L_\alpha (I_{\mrm{c}2(4)} - I_{\mrm{c}1(3)})/\Phi_0$, where each $I_{\mrm{c}j}$ corresponds to the critical current $I_{\mrm{c}}$ of the $j$th Josephson junction. 

Note that Eq. (\ref{eq:two bit potential without approximations}) contains four degrees of freedom, namely $\varphi_{i}$ and $\varphi_{i\mathrm{dc}}$, where $i = 1,2$. This type of compound SQUID is generally constructed with $\gamma_i \gg 1$ \cite{Han_Lapointe_Lukens_1992, Ray_Crutchfield_2023}. This implies that any changes in $\varphi_{i\mrm{xdc}}$ are rapidly observed in $\varphi_{i\mrm{dc}}$, so we assume $\vpdcIndexI = \vpxdcIndexI$. Applying this approximation to Eq. (\ref{eq:two bit potential without approximations}) projects the potential onto two dimensions, providing the potential energy surface shown in Fig. \ref{fig:landscape mem state instantiation} in the \vphiOnevphiTwoPlane.

Next, we assume that the fabrication consistency of the JJ elements is such that $\delta \beta_1 = \delta \beta_2 = 0$: This assumption is fairly common, as it streamlines the process of analyzing the device's equation of motion. In practice, $\delta \beta_1 \neq \delta \beta_2 \neq 0$, which results in a slight offset in the wells' locations \cite{Saira_2020}. This induced asymmetry in a real device can be at least partially compensated by external flux parameters if necessary. Along similar lines, we let $\beta \coloneqq \beta_1 = \beta_2$ as well as $I_{\mrm{c}} \coloneqq I_{\mrm{c}1} = I_{\mrm{c}2} = I_{\mrm{c}3} = I_{\mrm{c}4}$. Finally, assuming small-valued coupling ratios, we drop terms that are quadratic in $\mu$, which sends $\alpha \rightarrow 1$ and $L_\alpha \rightarrow L$.

All this done, rewriting Eq. (\ref{eq:two bit potential without approximations}) then gives:
\begin{align}
U' & = \dfrac{1}{2} (\varphi_1 -
\varphi_{1\mathrm{x}})^2 + \dfrac{1}{2} (\varphi_2 -
\varphi_{2\mathrm{x}})^2 \nonumber \\
&  \quad - \beta \cos \dfrac{\varphi_{1\mathrm{xdc}}}{2} \cos
\varphi_1 - \beta \cos \dfrac{\varphi_{2\mathrm{xdc}}}{2} \cos
\varphi_2  \nonumber \\
& \quad + \mu(\varphi_1 - \varphi_{1\mrm{x}})(\varphi_2 - \varphi_{2\mrm{x}}) ~.
\label{eq:two bit potential with useful approximations}
\end{align}
Equation (\ref{eq:two bit potential with useful approximations}) describes a dimensionless potential $U^\prime$ with two degrees of freedom; this will be frequently referred to in the remainder. We manipulate the potential energy landscape with the following circuit parameters for $i =1,2$: (i) $\varphi_{i\mathrm{x}}$ tilts the potential with respect to the $i$th axis, (ii) $\varphi_{i\mrm{xdc}}$ provides an in-situ barrier control on the $i$th axis, and (iii) $\mu$ supplies a coupling interaction between the two circuits, which biases the potential to favor wells that lie on one of the two diagonals in the \vphiOnevphiTwoPlane. 

\section{Control Erase Protocol Implementation}\label{sec:NANDGateDeviceImplementation}

The approximations in Sec. \ref{sec:DeviceImplementation} yield the potential given in Eq. (\ref{eq:two bit potential with useful approximations}) and displayed in Fig. \ref{fig:landscape mem state instantiation}. Manipulating the aforementioned external circuit parameters implements any of the eight possible control erase (CE) protocols shown in Fig. \ref{fig:AllControlErasureProtocols}. Using the CE $\CEOne$ from Figs. \ref{fig:erasure skeletons} and \ref{fig:CE example tableau} as an example, we detail how to find circuit parameter values that give rise to ``effective" protocols in Sec. \ref{sec:FindingEffectiveCEProtocols}. We qualitatively show this by exploring circuit parameter values that give rise to ranges for which this particular CE can be executed. Furthermore, we quantify this protocol's effectiveness---i.e., if it is highly robust---in Sec. \ref{sec:CERobustness}. Section \ref{sec:DetermineCEParameters} generalizes to determine which circuit parameters will be used to carry out any CE protocol in Fig. \ref{fig:AllControlErasureProtocols}. With this, Sec. \ref{sec:NANDGateApplication} demonstrates the physical execution of an effective NAND gate.
 
\subsection{Effective CE Protocol}\label{sec:FindingEffectiveCEProtocols}

\begin{figure*}[!t]
    \centering
    \begin{subfigure}
        \centering
        \begin{tikzpicture}
        \node[inner sep=0pt] (ce) at (0,0)
        {\includegraphics[width=0.85\linewidth]{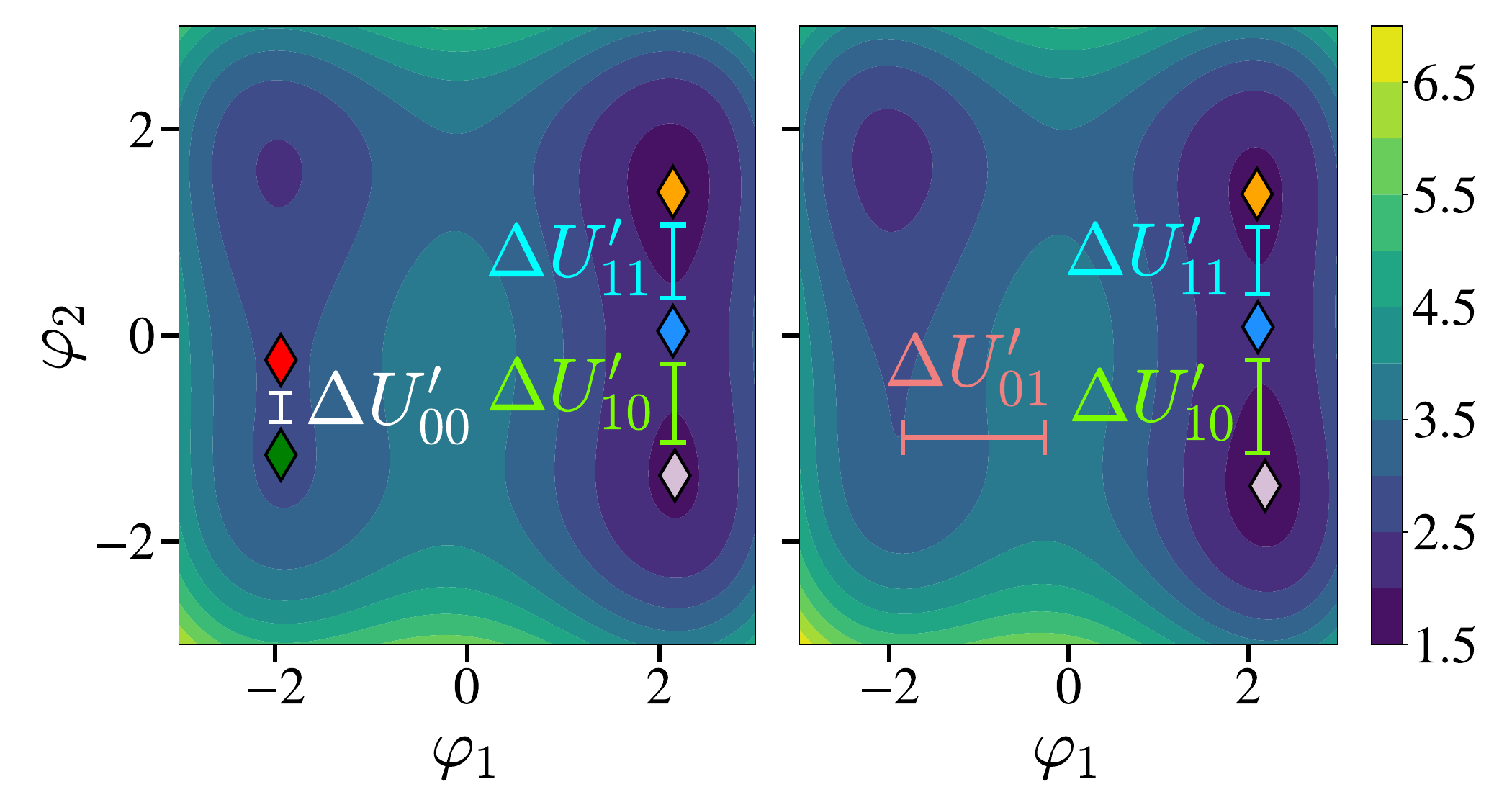}};
        \node[align=center] at (-7,3.37) {\large{\textbf{(a)}}};
        \end{tikzpicture}
    \end{subfigure}
    \begin{subfigure}
        \centering
        \begin{tikzpicture}
        \node[inner sep=0pt] (ce) at (0,0)
        {\includegraphics[width=0.85\linewidth]{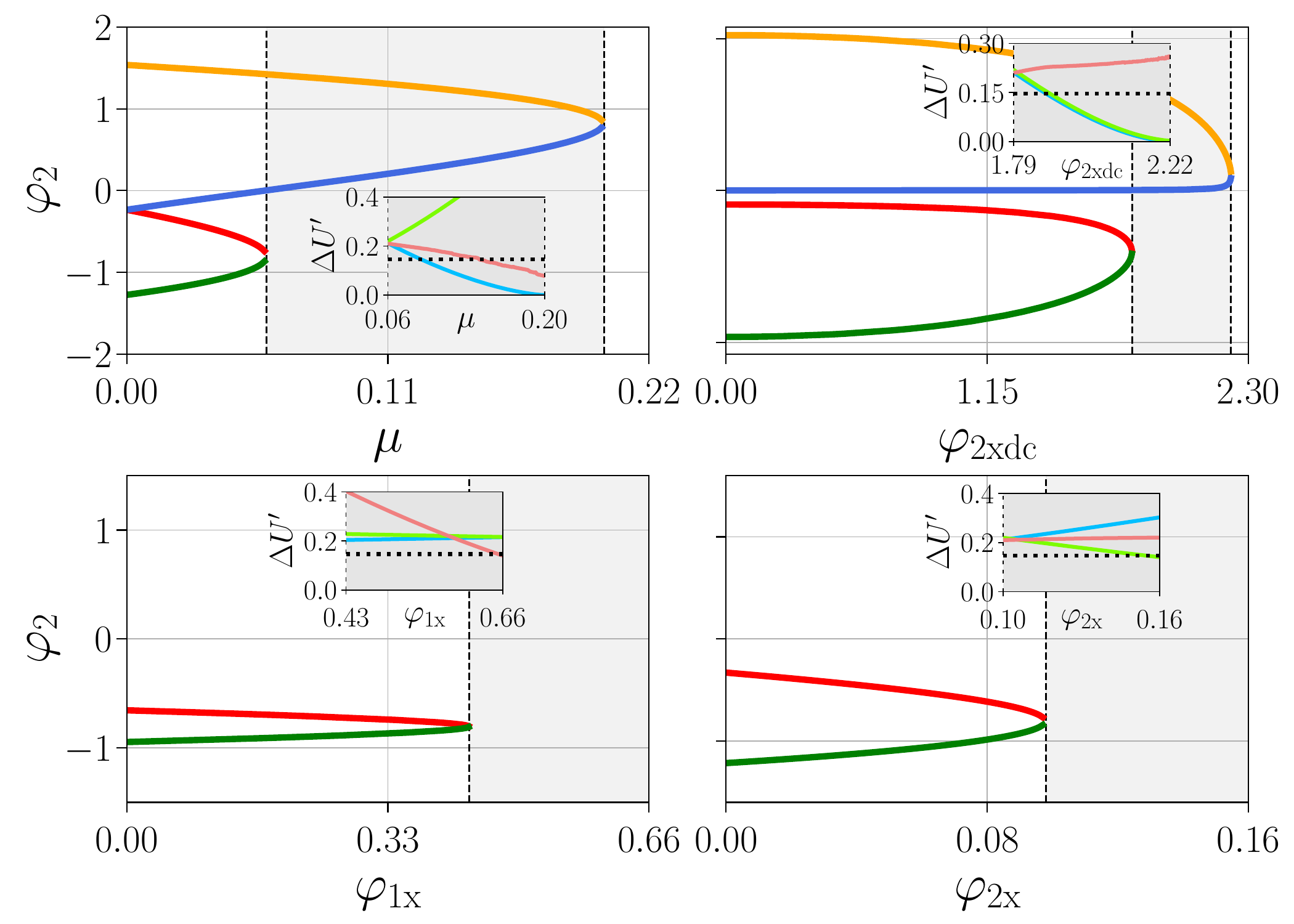}};
        \node[align=center] at (-7,4.75) {\large{\textbf{(b)}}};
        \end{tikzpicture}
    \end{subfigure}
    \caption{Circuit parameter ranges that eliminate the dimensionless potential barriers $\dUZeroZero$ and maintain $\dUOneOne$ and $\dUOneZero$. In the insets, light blue, light green, and light red correspond to the barriers $\dUOneOne$, $\dUOneZero$ and $\dUZeroOne$, respectively. The dotted line corresponds to a thermal energy barrier of $50 k_B T$ at $T = 100$ mK with $L = 230 \; \text{pH}$. For exemplary purposes, $\beta$ = 2.3. (a) Specific fixed points (colored diamonds) within the potential energy landscape, as well as corresponding energy barriers, are utilized to investigate the CE in Fig. \ref{fig:erasure skeletons}. (Left) Prior to the fixed point annihilation in the third quadrant. (Right) After fixed point annihilation that causes $\dUZeroZero$ to vanish. Once all information stored in the $00$ state is erased into the $01$ state and the potential is subsequently brought back to its original configuration, the CE is completed. (b) Circuit parameter ranges for achieving the CE in Fig. \ref{fig:erasure skeletons}, which can be accomplished by the annihilation of the green and red fixed points while maintaining the yellow and blue fixed points. Each circuit parameter range holds all other nonzero circuit parameters at the respective constant values: $\mu = 0.06$, $\vpxdcTwo = 1.79$, $\vpxOne = 0.61$, and $\vpxTwo = 0.10$. These circuit parameters lead to effective CE protocols and thereby robust NAND gates.}
    \label{fig:circuit parameter ranges}
    \label{}
\end{figure*}

Figure \figNumeric{\ref{fig:circuit parameter ranges}}{a} highlights the important fixed points on the potential immediately before and after the stable (red) and unstable (green) fixed points annihilate in the negative-$x$ half plane. After the annihilation, a CE is possible since the orange, blue, and purple fixed points on the positive-$x$ half plane are preserved. From the perspective detailed in Sec. \ref{sec:ControlErasureProtocolSetup}, the information stored in the $\tY$ state on the negative-$x$ half plane is erased, while all other information in the potential's memory states is preserved. In this example, $\dUZeroZero$ denotes the potential barrier that must vanish for the CE to be carried out. Once this barrier vanishes, $\dUZeroOne$, $\dUOneZero$, and $\dUOneOne$ denote the barriers that need to be maintained. The goal is to find the circuit parameter values that give these barriers sufficiently large heights, as this further contributes to a robust CE. 

Figure \figNumeric{\ref{fig:circuit parameter ranges}}{b} illustrates bifurcation diagrams of the $\varphi_2$ coordinate over a given circuit parameter range, while all other parameters are held constant. Note that the coordinate $\varphi_2$ is solely illustrated because $\varphi_1$ varies negligibly in comparison. Additionally, if fixed points annihilate within a parameter value range, they are displayed in Fig. \figNumericNoSpace{\ref{fig:circuit parameter ranges}}{b}. Otherwise, the fixed points do not annihilate within this selected range, and their $\varphi_2$ trajectory will not be shown.

To guide the exploration of viable parameter values, a general view of the relationship between the dimensionless barrier heights illustrated in Fig. \figNumeric{\ref{fig:circuit parameter ranges}}{b} and the energy scale of thermal fluctuations $k_B T$ is useful. Understanding this aids in determining if a barrier height is energetically large enough to store microstates within a metastable memory state. 

For example, suppose the barrier should be no smaller than $50 k_B T$. Using common values from the SQUID literature \cite{Han_1992_rf, Ozfidan_2020}, $T = 100 \; \text{mK}$ and $L = 230 \; \text{pH}$, this corresponds to a dimensionless barrier height of $\Delta U^\prime = 50 k_B T / U_0 = 0.15$. We indicate this barrier height with the dashed horizontal line in the insets of Fig. \figNumericNoSpace{\ref{fig:circuit parameter ranges}}{b}; it represents a proxy for characterizing dimensionless barrier heights. That is, if a circuit parameter's value results in either $\dUZeroOne$, $\dUOneZero$, or $\dUOneOne$ falling below this line, then we say it is no longer able to reliably store information in the relevant well. Of course, this all varies depending on the device's operating temperature and parameter values. For instance, if we wanted to continue using $U^{\prime} = 0.15$ as a standard value from the QFP literature \cite{Hosoya_Goto_1991, Takeuchi_Ozawa_Yamanashi_Yoshikawa_2013, Takeuchi_Yamanashi_Yoshikawa_2015, Takeuchi_2022}---that being $T = 4.2 \; \text{K}$ and $L = 10 \; \text{pH}$---this corresponds to a dimension-full barrier height of $\Delta U = 0.15 U_0 = 28 k_B T$. As a final note, motivated by Ozfidan et al \cite{Ozfidan_2020}, we let $\beta = 2.3$.

The following now describes the qualitative consequences of changing each circuit parameter and the resulting potential in Fig. \figNumericNoSpace{\ref{fig:circuit parameter ranges}}{a}. First, as $\mu$ increases, the depth of the minima representing the states $01$ and $10$ increases. Meanwhile, the minima's depth in the $11$ state decrease, leading to $\dUOneOne$ decreasing. This indicates that $\mu$ should not take on too large of a value in order to avoid fluctuations over the $\dUOneOne$ barrier. Next, if the magnitude of $\vpxdcTwo$ increases, then both $\dUOneOne$ and $\dUOneZero$  decrease. Consequently, its value should be large enough only to ensure that $\dUZeroZero$ is eliminated while $\dUOneOne$ and $\dUOneZero$ are maintained. Changing $\vpxOne$ adjusts the potential's tilt with respect to the $\varphi_1$ axis. If $\vpxOne$ takes on too large of a value, then the barrier $\dUZeroOne$ will be undesirably eliminated. However, its value must be large enough to ensure that $\dUZeroZero$ is eliminated. Similarly, tuning $\vpxTwo$ changes the tilt of the potential with respect to the $\varphi_2$ axis. If $\vpxTwo$ is too large, then the blue (purple) stable (unstable) fixed points can annihilate, which causes information to be erased in the region where it should be preserved. 

Obtaining effective circuit parameter ranges involves determining two particular values: (i) the value resulting in $\dUZeroOne$, $\dUOneZero$, and $\dUOneOne$ having the same height, and (ii) the value which results in one barrier falling below the proxy line. The first (second) criterion  serves as the lower (upper) end of the range. Ultimately, there are clear tradeoffs that depend on the value of a given circuit parameter. Finding the set of parameters permitting control over which information is erased and preserved requires a balancing act: A careful process of selecting parameter values that are large enough to erase the desired information, but not so large that the information originally planned to be preserved ends up being erased. 

For $\mu$, since $\dUOneOne$ is most susceptible to information leakage, its effective range corresponds to $[0.06, 0.09]$. Similarly, to maintain as large $\dUOneOne$ and $\dUOneZero$ as possible, the value of $\vpxdcTwo$ should also lie at the lower end of the range $[1.79, 1.88]$. Then, to aid the effect of $\mu$ and $\vpxdcTwo$ on the potential and to maintain $\dUZeroOne$, $\vpxOne$ would ideally take on a value within the range of $[0.59, 0.65]$. At the same time, to strike a balance between the increase of $\dUOneOne$ and the decrease of $\dUOneZero$, $\vpxTwo$ ideally lies within the range $[0.1, 0.15]$. Within these ranges, a reliable CE can be performed.

\subsection{Control Erase Parameter Selection}
\label{sec:DetermineCEParameters}

We now detail which circuit parameters---including their respective signs and magnitudes---are deployed for any of the eight CE protocols of interest. By leveraging the relationship between the potential's dynamics and employing the CE notation introduced in Sec. \ref{sec:ControlErasureProtocolSetup}, all eight CE protocols in Fig. \ref{fig:AllControlErasureProtocols} can be implemented in a systematic way. The following explains this approach, summarizing the results in Table \ref{tab:table of CEs and notation}.

To begin, define a sign-based Iverson bracket:
\begin{equation}
    \ModIversonBracketNotation \coloneqq \begin{cases}
        > 0 & \text{if} \ [P] = 1 ~,\\
        < 0 & \text{if} \ [P] = 0 ~.
    \end{cases}
\end{equation}
Here, $[P]$ is the Iverson bracket of the statement $P$, and $[P]$ evaluates to 1 if $P$ is true or to $0$ if $P$ is false. This notation is useful when discussing how the sign of a circuit parameter is determined. Let's now review the four parameters involved in each CE protocol:
\begin{itemize}
    \item $\vpxdcIndexI$ lowers the barrier between specific pairs of states to be erased;
    \item $\vpxBarrierOffset$ compensates for the unwanted effects of $\vpxdcIndexI$;
    \item $\mu$ biases the target erasure state to be more energetically favorable; and
    \item $\vpxBiasOffset$ compensates for the unwanted effects of $\mu$.
\end{itemize}
For these parameters, $i, j \in \{1,2\}$ with $i \neq j$. To determine which $\vpxdcIndexI$ to use for a given CE, recall that we specify a CE with three parameters: $\rho$, $\sigma$, and $A$. If $\rho = \tX \; (\tY)$, then $i,j = 2,1 \; (1,2)$. As seen from Eq. (\ref{eq:two bit potential with useful approximations}), $\vpxdcIndexI$ appears only in the $\mathrm{cosine}$ function: Since this function is even, the sign of $\vpxdcIndexI$ is not relevant. Recall from Sections \ref{sec:ControlErasureProtocolSetup} and \ref{sec:FindingEffectiveCEProtocols} that $\vpxdcIndexI$ lowers energy barriers on both sides of the $\vpxdcIndexOnly$th axis. However, the CE requires one barrier to be maintained. Thus, we need a control parameter to offset this barrier drop so that it applies to only one half of the \vphiOnevphiTwoPlane, while maintaining the other barrier. The parameter deploying this barrier offset is $\vpxBarrierOffset$, and its sign is given by $\ModIversonBracket{\sigma \neq \rho}$. 

Next, $\mu$ biases the potential such that the minima along one of the diagonals of the \vphiOnevphiTwoPlane $\ $are more energetically favorable compared to those on the other diagonal. Its sign determines which well becomes the target for the erasure and is given by $\ModIversonBracket{[\sigma = \rho]  \ \mathrm{XOR} \ A}$. Since $\mu$ acts along the diagonals of the \vphiOnevphiTwoPlane, it causes asymmetry between two wells that are supposed to be storing information, resulting unwanted information leakage. Fortunately, the parameter $\vpxBiasOffset$ can be used to compensate this effect by tilting the potential to offset this unintentional information leakage. The sign of this parameter is found through $\ModIversonBracket{A}$. 

For all CEs of interest in Fig. \ref{fig:AllControlErasureProtocols}, Table \ref{tab:table of CEs and notation} compactly shows the signs of the relevant circuit parameters.

\subsection{CE Robustness}
\label{sec:CERobustness}

During our example CE in Fig. \ref{fig:circuit parameter ranges}, once $\dUZeroZero$ vanishes, the information needs to propagate from the $00$ state to the $01$ state. The time this takes is denoted $\tCE$; it serves as a proxy for the CE timescale.

As this information is erased, other information is being stored in the $01$, $10$, and $11$ states due to the barriers $\dUZeroOne$, $\dUOneZero$, and $\dUOneOne$ being maintained. However, we see in Fig. \figNumeric{\ref{fig:circuit parameter ranges}}{b} that as a circuit parameter increases, at least one barrier's height decreases. The microstates stored in these decreasing barriers are most susceptible to fluctuating out of their respective memory states. Due to this, the timescale that characterizes how long these states reliably store information is in competition with $\tCE$. This type of timescale is known as a dwell time \cite{Han_Lapointe_Lukens_1989, Han_Lapointe_Lukens_1992, Hanasaki_Nemoto_Tanaka_2019}, which is denoted $\tDwell$. Implicitly from Fig. \figNumericNoSpace{\ref{fig:circuit parameter ranges}}{a}, there are three dwell times that contribute to CE performance---those being the dwell times corresponding to the states $01$, $10$, and $11$. The minimum of these three dwell times ultimately dictates CE failure.

If the length of time that information can be reliably stored is much longer than the timescale of a CE protocol, then the probability of an error resulting from an unwanted thermally activated transition is low. To quantify this, we take the ratio $\tDwell/\tCE$ as a measure of robustness.

To calculate $\tCE$, we approximate the information in the $00$ state as a single damped particle travelling down a quadratic potential from the point where the fixed points initially annihilate; the derivation is included in App. \ref{app:LinearQuadraticApproximations}. For $\tDwell$, since there are three dwell times that affect CE performance, we choose the minimum of these dwell times to characterize CE robustness. This provides a more conservative estimate of a particular CE's performance; refer to App. \ref{app:DwellTimes} for details. Of particular relevance, $\tDwell$ is exponentially proportional to $1/L$ and $1/T$; this has a dominant effect on the robustness $\tDwell/\tCE$. Studying the robustness as a function of $L$ and $T$ allows for interpreting the performance of a CE across the MQP and QFP literature.

\begin{figure*}[!t]
    \centering
    \includegraphics[width=\linewidth]{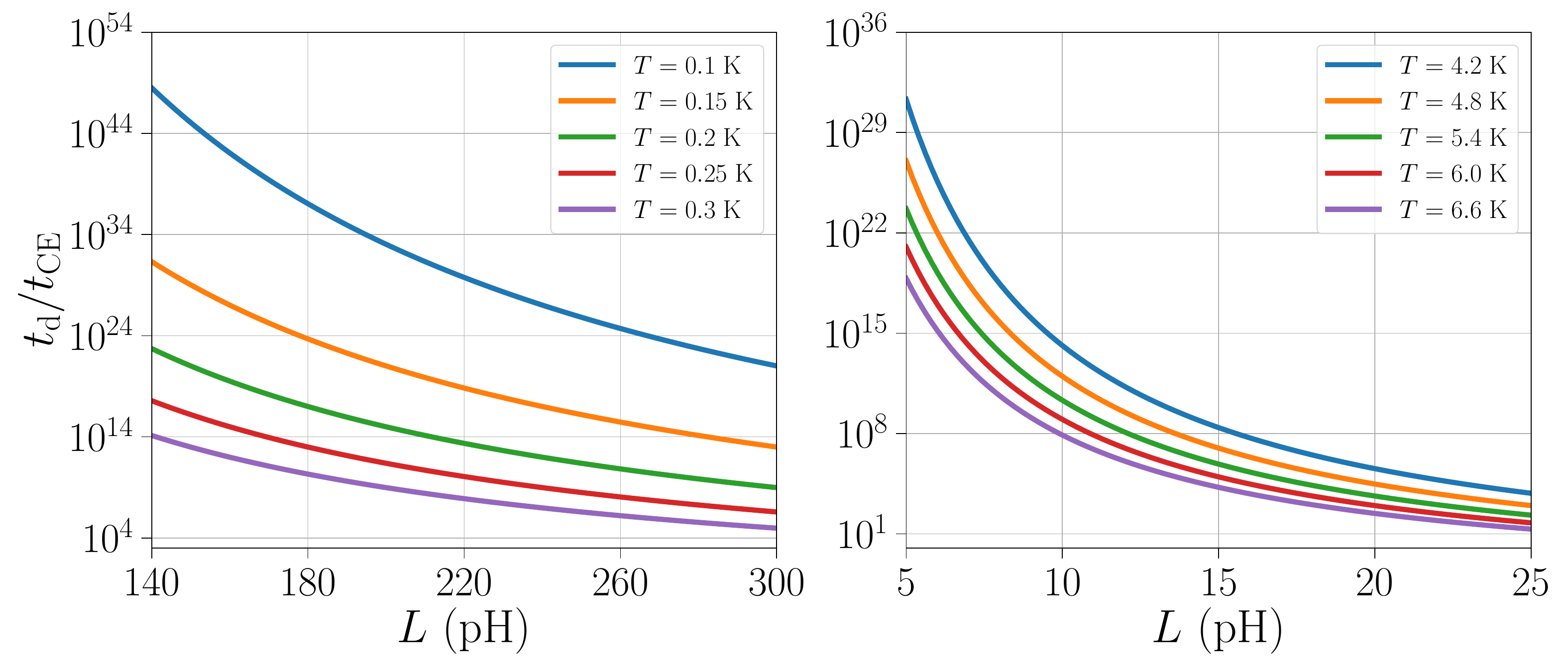}
    \caption{The ratio $\tDwell/\tCE$ characterizes the robustness of a particular device's construction as an order of magnitude estimate. (Left) MQP literature inductance values and temperatures \cite{Berkley_Johnson_2010, Han_1992_rf, Han_Lapointe_Lukens_1989, Han_Lapointe_Lukens_1992, Harris_Berkley_Johnson_2007, Harris_Johnson_Han_2008, Harris_Lanting_2009, Saira_2020, Ray_Crutchfield_2023, Wimsatt_Saira_Boyd_Matheny_Han_Roukes_Crutchfield_2021}. To demonstrate the behavior of the robustness at colder temperatures, the temperature range $[100-300] \; \text{mK}$ is only displayed. Note that $R = 1 \; \Omega$, $C = 500 \; \text{fF}$, and $I_c = 5 \; \mu \text{A}$. (Right) QFP literature inductance and temperature values \cite{Hosoya_Goto_1991, Takeuchi_2022, Takeuchi_Ozawa_Yamanashi_Yoshikawa_2013, Takeuchi_Yamanashi_Yoshikawa_2015}. Note that the values $5 \leq L \leq 10 \; \text{pH}$ are exercised to demonstrate the performance of a QFP if obtainable, while typical QFP inductances are greater than or equal to $10 \; \text{pH}$. For this figure, $R = 1 \; \Omega$, $C = 1 \; \text{pF}$ and $I_c = 25 \; \mu \text{A}$.}
    \label{fig:robustness}
\end{figure*}

From Sec. \ref{sec:FindingEffectiveCEProtocols}, selecting parameter values at the lower end of the effective parameter range yields a more robust CE: We found that $\mu = 0.06$, $\vpxdcIndexI = 1.79$, $\vpxBarrierOffset = 0.61$, and $\vpxBiasOffset = 0.10$. These will be known as the \emph{effective parameter values}. With these, Fig. \ref{fig:robustness} displays the CE robustness for different values of $L$ and $T$. Since the MQP (QFP) literatures employ SQUIDs for different purposes---thereby leading to differing ranges of $L$ and $T$---we separate the possible value ranges into the left (right) figures. The relevant MQP literature \cite{Berkley_Johnson_2010, Han_1992_rf, Han_Lapointe_Lukens_1989, Han_Lapointe_Lukens_1992, Harris_Berkley_Johnson_2007, Harris_Johnson_Han_2008, Harris_Lanting_2009, Saira_2020, Ray_Crutchfield_2023, Wimsatt_Saira_Boyd_Matheny_Han_Roukes_Crutchfield_2021} corresponds to the left figure, which has a general temperature range of $[100, 500] \; \text{mK}$, while the inductance values range from $[140, 300] \; \text{pH}$. An upper value of $\mathcal{O}(10^{48})$ CEs per device is obtained with the inductance value of $L= 140 \; \text{pH}$ found from Saira et al \cite{Saira_2020} at $T = 100 \; \text{mK}$. 

On the right---following the QFP literature \cite{Hosoya_Goto_1991, Takeuchi_2022, Takeuchi_Ozawa_Yamanashi_Yoshikawa_2013, Takeuchi_Yamanashi_Yoshikawa_2015}---the operating temperatures are no colder than $4.2 \; \text{K}$, in addition to $L \geq 10 \; \text{pH}$. To demonstrate robustness with a broader scope than constructed thus far, we use a range that begins at a theoretical value of $L = 5 \; \text{pH}$ to obtain an upper value of $\mathcal{O}(10^{31})$ CEs per device at $4.2 \; \text{K}$. For both areas of literature, we identify that lower temperature and inductance values correspond to a greater expected number of successful CEs per device. The values of $R$, $C$, and $I_c$ are chosen as typical values found in the respective literatures.
As a final point, note that from Fig. \ref{fig:robustness} the ratio $\tDwell / \tCE$ was calculated with the effective parameter values. We further assumed that the information travelling from the $00$ state to the $01$ state begins at the location determined from nonzero circuit parameters for the CE. 

Now, suppose all circuit parameters were initialized to zero, so that the potential matches Fig. \ref{fig:landscape mem state instantiation}. Then, tune all circuit parameters infinitely fast to their respective final values. Due to this, the information stored in the $00$ state is now travelling from the location determined by zero-valued circuit parameters, as opposed to the point of annihilation. This means that the information must now travel a greater distance to the $01$ state, thereby resulting in a larger $\tCE$. Considering this scenario provides a more conservative estimate for the ratio $\tDwell / \tCE$. That said, we found that this larger $\tCE$ will be no more than twice the value of the most conservative estimate of $\tCE$, leading to a reduction of the robustness of order one compared to that shown in Fig. \ref{fig:robustness}.

\subsection{NAND Gate Application}
\label{sec:NANDGateApplication}

\begin{figure*}[!t]
    \centering
    \includegraphics[width=\linewidth]{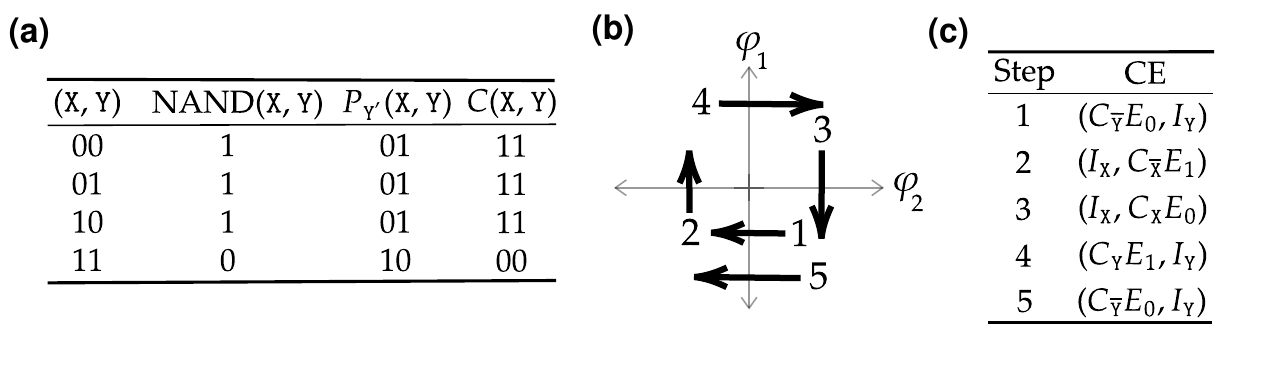}
    \caption{NAND computations carried out on a coarse-grained potential of Fig. \ref{fig:landscape mem state instantiation}, which is reproducible by Eq. (\ref{eq:two bit potential with useful approximations}). (a) Tabulated computations of the NAND gate. Steps (1)-(3) execute the partial computation $\PartialCompExample{\tYprime}$. Additionally, steps (1)-(5) accomplish the complete computation $C[(\tX, \tY)]$. (b) Arrow-based schematic for performing partial and complete NAND gates. (c) Tabulated step numbers and CE protocols corresponding to Fig. (b).}
    \label{fig:NAND computations with arrows}
\end{figure*}

Executing successive CE protocols allows for performing a NAND gate. The CE has two input bits and two output bits as seen from Fig. \ref{fig:CE example tableau}, while the NAND gate has two inputs but only one output bit. When the number of logical output bits is less than the number of embedded output bits in a computational system, there is added flexibility when choosing which bit to read as the output for a computation. For example, we can specify $\tXprime = \text{NAND}[(\tX,\tY)]$ without prescribing the $\tYprime$ computation. Thus, we define two kinds of NAND gates: a \textit{partial} NAND, written as $\PartialCompNotation$, which only requires the state $\rho^\prime$ to yield the desired output, and a \textit{complete} NAND, denoted as $\CompleteCompNotation$, which requires that $\tXprime = \tYprime = \text{NAND}[(\tX,\tY)]$. Fig. \figNumeric{\ref{fig:NAND computations with arrows}}{a} tabulates these choices.

From here, Fig. \figNumeric{\ref{fig:NAND computations with arrows}}{b} illustrates both kinds of computations, and Fig. \figNumeric{\ref{fig:NAND computations with arrows}}{c} tabulates the computations. Recall from Fig. \ref{fig:landscape mem state instantiation} that the first, second, third, and fourth quadrants correspond to the memory states $11$, $01$, $00$, and $10$, respectively. In this example, carrying out only steps (1)-(3) executes the partial computation $P_{\tYprime}$. This can seen by tracking the arrows through the included protocol steps: In step (1), the information in the state $10$ is erased into the $00$ state; in step (2), the combined information---comprising of the information from the initial $00$ and $10$ states---is erased into the $01$ state; in step (3), the information in the $11$ state is translated into the $10$ state using the dynamics of the CE protocol. Note that in this step there is no logical erasure since quadrant III was already vacated in step (1). At this point, the $\tYprime$ output bit represents a NAND gate of $\tX$ and $\tY$. Continuing to track the arrows for steps (4)-(5) executes the computation $C$.

Partial NAND gates built from CEs require fewer successive CE protocol executions than complete NAND gates. However, performing complete NAND gates provides more redundancy regarding from which states the computations can be read.

\section{Conclusion}\label{sec:Conclusion} 

We presented the family of control erase (CE) protocols in Fig. \ref{fig:AllControlErasureProtocols} and introduced notation in Sec. \ref{sec:ControlErasureProtocolSetup} that allows them to be concisely enumerated. We introduced a device in Fig. \ref{fig:universal computing circuit} that performs CE protocols by tuning circuit parameter values to access desired bifurcations. Characterizing a measure of a protocol's robustness involved associating a storage length time with the protocol's duration. We conducted an order of magnitude estimate of the robustness for an example CE protocol in Fig. \ref{fig:robustness}, yielding a high robustness for protocols that employ realistic parameter values spanning different areas of superconducting circuit literature.

Section \ref{sec:DetermineCEParameters} established a connection between the device's circuit parameters and the CE notation. Linking various CE protocols together by leveraging this relationship allowed for universal computations. Section \ref{sec:NANDGateApplication} demonstrated a NAND gate that employs only CEs. Due to its superconducting operation, the device can perform CE-based universal computations at extremely low energy cost.

In addition, there is added flexibility when performing NAND gates via the potential in Fig. \ref{fig:landscape mem state instantiation}. Partial computations permit fewer executions of erasure protocols, which implies that computations can be performed with even lower energetic costs. Conversely, executing complete computations gives rise to reading the same computation regardless of which axis is chosen to be the output, thereby providing an added layer of redundancy. 

Follow-on efforts explore the thermodynamic efficiency of this NAND gate, run SPICE simulations of the protocol for varying circuit constructions, and investigate implementing other useful and more complex logical operations.

\section{Acknowledgements}

The authors thank Fabio Anza, Scott Habermehl, Jacob Hastings, Kuen Wai Tang, and Gregory Wimsatt for helpful comments and discussions, as well as the Telluride Science Research Center for its hospitality during visits, and the participants of the Information Engines workshop there for their valuable feedback. This material is based upon work supported by, or in part by, U.S. Army Research Laboratory, U.S. Army Research Office grant W911NF-21-1-0048.

\bigskip

\bibliography{main-bib}

\appendix

\section{Control Erase Timescale Approximation Scheme}\label{app:LinearQuadraticApproximations}

The following details an approximation scheme for the potential in Eq. (\ref{eq:two bit potential with useful approximations}) during the control erase (CE) protocol described in Sec \ref{sec:ControlErasureProtocolSetup}. Specifically, to represent the distribution of microstates being erased into the $01$ state, we approximate this situation as a particle sliding down a quadratic slope. From this, we calculate $\tCE$---the time it takes for the particle to travel to the bottom of the slope.

First, note that in Fig. \ref{fig:CE example tableau}, the coordinate $\phi_1$ has negligible change in comparison to the change in $\phi_2$. With this in mind, we will consider the potential to be dependent only on $\phi_2$. Next, the height of the potential $U$ that the particle travels down is on the order of $U_0 \sim 10^{-22}$: Essentially, the particle's motion can be described in terms of $\phi_2$ without loss of generality. Then, there are two forces acting on the particle: (i) the force due to the potential $-\mathrm{d}U/\mathrm{d}\phi_2$ and (ii) the classical damping force $-\gamma \mrm{d}\phi_2/\mrm{d}t$. This means that the particle's motion is described by Newtonian mechanics due to the influence of a dimension-full potential $U$ and damping coefficient $\gamma$, written as:
\begin{equation}\label{eq:AppA f=ma}
     \dfrac{\mrm{d}^2\phi_2}{\mrm{d}t^2} + \dfrac{\gamma}{m} \dfrac{\mrm{d}\phi_2}{\mrm{d}t} + \dfrac{1} {m}\dfrac{\mathrm{d}U}{\mathrm{d}\phi_2} = 0 ~.
\end{equation}
Observe that Eq. (\ref{eq:two bit potential with useful approximations}) is dimensionless, while Eq. (\ref{eq:AppA f=ma}) describes a particle's motion under the influence of a potential with units of energy. To reconcile this, we must rewrite Eq. (\ref{eq:AppA f=ma}) in terms of dimensionless quantities. Reference \cite{Ray_Crutchfield_2023} detailed a strategy for converting dimension-full equations of motion---such as Eq. (\ref{eq:AppA f=ma})---into dimensionless equations of motion. To do this, we write a dimensional quantity $z$ as $z = z'z_c$ in which $z'$ is the dimensionless representation of the quantity, while $z_c$ represents a dimensional scaling factor. Table \ref{tab:circuit quantity scalings} summarizes the circuit quantity scaling parameters $z_c$.

Now, after appropriately making substitutions and further simplifying, we write the dimensionless Eq. (\ref{eq:AppA f=ma}) as:
\begin{equation}
    \dfrac{\mrm{d}^2\varphi_2}{\mrm{d}t'^2} + \Lambda \dfrac{\mrm{d}\varphi_2}{\mrm{d}t'} + \theta \dfrac{\mrm{d}U'}{\mrm{d}\varphi_2} = 0 ~,
\end{equation}
where $\Lambda = \gamma' \sqrt{LC}/m' RC$ and $\theta = 1/m'$. 
\begin{table}[!t]
    \centering
    \begin{tabular}{cc}
    \toprule
         Circuit quantity & $z'(z_c)$ Identifications \\[0.5pt]
    \midrule
         $m$ & $m'(C)$ \\[2pt]
         $\gamma$ & $\gamma' (1/R)$ \\[2pt]
         $\phi$ & $\varphi (\phi_c)$ \\[2pt]
         $U$ & $U' (\phi_c^2/L)$ \\[2pt]
         $t$ & $t'(\sqrt{LC})$ \\[2pt]
    \bottomrule
    \end{tabular}
    \caption{Relevant circuit quantity scalings.}
    \label{tab:circuit quantity scalings}
\end{table}
The approximation scheme of the potential will now be detailed. First, we write the dimension-full potential as $U(\phi_2) = k\phi_2^2/2$, where $k$ serves as the spring constant of the dimension-full potential. The goal is to write the potential in the dimensionless form: $U'(\varphi_2) = k' \varphi_2^2/2$, with $k'$ being the spring constant of the dimensionless potential. This can be done by setting $k_c = U_c/2\phi_c^2$, in order to obtain $k' = U' / (\varphi_2)^2$.
Now, the particle's motion is written as:
\begin{align}\label{eq:initial f=ma}
    \dfrac{\mrm{d}^2 \varphi_2}{\mrm{d}t'^2} + 2\lambda \dfrac{\mrm{d}\varphi_2}{\mrm{d}t'} + \omega^2 \varphi_2 = 0 ~,
\end{align}
which is analogous to the differential equation modelling a damped harmonic oscillator that oscillates at frequency $\omega^2 = k'\theta$ under damping coefficient $\lambda = \Lambda/2$. We deploy the ansatz: $\varphi_2(t') = A\mrm{e}^{\alpha t'}$ in which $\alpha = -\lambda \pm \sqrt{\lambda^2 - \omega^2} = -\lambda \pm \Omega$. We will focus on two possible situations which involve $\Omega^2 > 0$ or $\Omega^2 < 0$. Subsequently, these will be categorized as \emph{overdamped} and \emph{underdamped}, respectively.

\subsubsection{Overdamped}

When $\Omega^2 > 0$, the solution to Eq. (\ref{eq:initial f=ma}) is \cite{Morin_2012}:
\begin{equation}\label{eq:AppA OverdampedInitial}
    \varphi_2(t') = A \exp(-(\lambda - \Omega)t') + B \exp(-(\lambda + \Omega)t') ~,
\end{equation}
which describes an overdamped oscillator with initial conditions $A$ and $B$. To understand them, we first detail the consequences of the particle-like representation of the distribution of microstates. The average position of this distribution localizes to a position at any given time---represented by a particle in this case---while its average initial velocity will be zero. With this, the initial conditions are (i) $\varphi_2(t'=0) = \initPos$ and (ii) $\mathrm{d}\varphi_2(t'=0)/\mrm{d}t' = 0$. Solving for these two initial conditions, and substituting the results back into Eq. (\ref{eq:AppA OverdampedInitial}), gives:
\begin{equation}\label{eq:AppA finalOverdamped}
    \varphi_2(t') = \dfrac{\initPos}{\Omega} (\lambda + \Omega) \mrm{e}^{-\lambda t'} \sinh (\Omega t') + \initPos \mrm{e}^{-(\lambda + \Omega)t'} ~.
\end{equation}
Numerically solving Eq (\ref{eq:AppA finalOverdamped}) for a final time $t' = t_f'$ and multiplying the result by $\sqrt{LC}$, yields $\tCE = t_f' \sqrt{LC}$.

\subsubsection{Underdamped}

If $\Omega^2 < 0$, we must define $\widetilde{\omega} = \sqrt{\omega^2 - \lambda^2}$ in order for the general solution to be written as \cite{Morin_2012}:
\begin{equation}\label{eq:AppA quad appx original phi_2 solved equation}
    \varphi_2(t') = D \exp(-\lambda t') \cos(\widetilde{\omega}t' + \theta) ~.
\end{equation}
Equation (\ref{eq:AppA quad appx original phi_2 solved equation}) describes an underdamped harmonic oscillator for which the coefficient $D$ and phase $\theta$ are solved using initial conditions. Subsequently, solving for them in the same manner as before and substituting the results into  Eq. (\ref{eq:AppA quad appx original phi_2 solved equation}) produces:
\begin{align}
    \varphi_2(t') &= \initPos \sqrt{1 + \left(\dfrac{\lambda}{\tildeOmega}\right)^2} \exp(-\lambda t') \cos(\tildeOmega t' + \Theta) \nonumber \label{eq:final phi_2 without approximations} ~,
\end{align}
for which $\Theta = \arctan(\lambda / \widetilde{\omega})$. 

\section{Dwell Time}\label{app:DwellTimes}

The escape rate \cite{Hanggi_1986, Sharifi_1988, Saira_2020} in the $\EscapeRateIndexOnly = 1,2$ direction is:
\begin{equation}\label{eq:escape rate definition}
    \GammaIndexed = \dfrac{\PlasmaFreqIndexed}{2\pi}\exp\left(-\dfrac{\Delta U_b}{k_B T}\right) ~.
\end{equation}
First, $\PlasmaFreqIndexed$ is the plasma frequency in the $\EscapeRateIndexOnly$th flux direction. It serves as the characteristic frequency of oscillations that are parallel to the escape direction; i.e., how frequently the particle approaches the barrier. To find it, we consider the dimension-full potential $U(\phi_1, \phi_2)$ and identify that we can perform a Taylor expansion in the $\EscapeRateIndexOnly$th coordinate around a minimum $(\phi_1 = \pOneMin, \phi_2 = \pTwoMin)$. Of particular interest, and up to a constant, we look to the following second order terms of the potential:
\begin{equation}
    U(\phi_1, \phi_2) \approx \dfrac{1}{2} \dfrac{\partial^2U} {\partial \phi_1^2}\bigg\vert_{(\pOneMin, \pTwoMin)} (\phi_1 - \pOneMin)^2 ~,\\[0.5pt]
\end{equation}
as well as:
\begin{equation}
    U(\phi_1, \phi_2) \approx \dfrac{1}{2} \dfrac{\partial^2 U} {\partial \phi_2^2}\bigg\vert_{(\pOneMin, \pTwoMin)} (\phi_2 - \pTwoMin)^2 ~.
\end{equation}
We identify the second partial derivative taken with respect to $\phi_l$ evaluated at the minimum to be the spring constant $k_\EscapeRateIndexOnly$. From here, we take the corresponding plasma frequency to be $\PlasmaFreqIndexed = \sqrt{k_\EscapeRateIndexOnly/m}$ for which the dimension-full mass $m = C$.

Furthermore, $\Delta U_b$ indicates the barrier height between a local minimum and a local maximum, $k_B$ is the Boltzmann constant, and $T$ is the circuit's operating temperature. The escape rate is exponentially damped by the energy barrier because the thermal environment guarantees particles are exponentially unliklely to have energies that exceed the thermal energy scale. $\GammaIndexed$ characterizes how many escape events are expected out of the minima per second. Finally, the dwell time is the reciprocal of Eq. (\ref{eq:escape rate definition}):  $\tDwellIndexed = 1/\GammaIndexed$. Explicitly, the plasma frequencies of interest for the dimensionless barriers $\dUZeroOne$ ($\dUOneOne$ and $\dUOneZero$) correspond to the flux direction $l = 1$ ($l = 2$). We utilize the minimum dwell time to calculate the robustness $\tDwell/\tCE$.

\section{Control Erasures of Interest}\label{app:AllControlErasureProtocols}
\label{app:CEs}

\begin{figure*}[t]
    \centering
    \includegraphics[width=\linewidth]{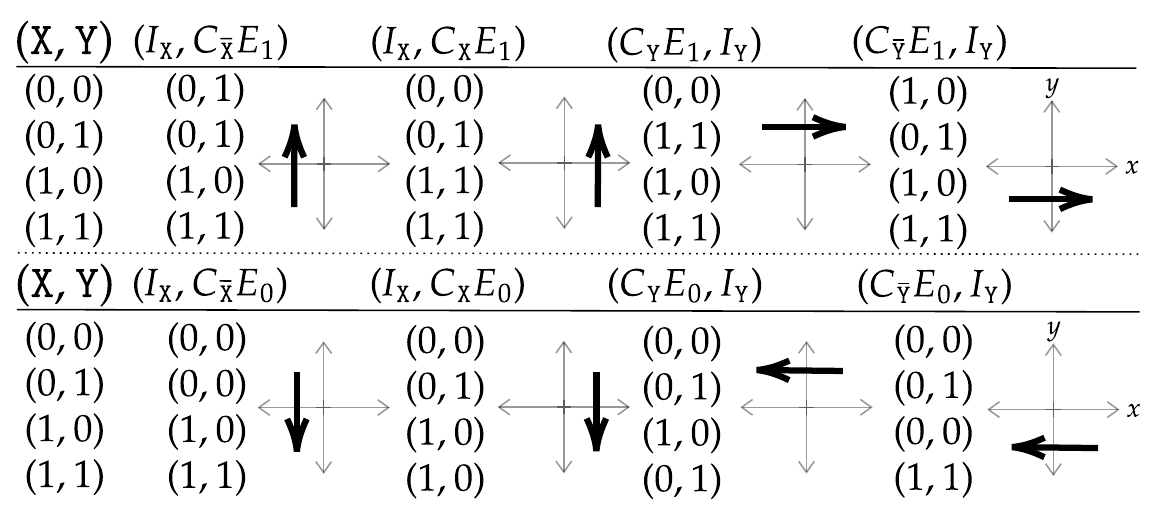}
    \caption{All control erasure protocols of interest here, displayed with arrow-based notation first shown in Fig. \ref{fig:CE example tableau}.}
    \label{fig:AllControlErasureProtocols}
\end{figure*}

Figure \ref{fig:AllControlErasureProtocols} exhibits all control erasure protocols of interest. Appropriate successive executions of a subset of these protocols lead to performing a NAND gate. Note that this is not an exhaustive illustration of possible control erasure protocols, but only represents those of current interest. Table \ref{tab:table of CEs and notation} shows the indices, signs and relationships of the relevant circuit parameters for a particular CE. The nonzero barrier-control parameter is determined by $i$. Meanwhile, the magnitudes of the tilt parameters that offset the bias and barrier changes to the potential are related by $|\vpxBarrierOffset| > |\vpxBiasOffset|$, respectively. By using the effective magnitudes of the circuit parameters detailed in Sec. \ref{sec:CERobustness} and successively executing effective CE protocols, a robust NAND gate can be performed.

\begin{table*}[t]
    \centering
    \begin{tabular}{ccccccccc}
        \toprule
          & $\CEOne$ & $\CETwo$ & $\CEThree$ & $\CEFour$ & $\CEFive$ & $\CESix$ & $\CESeven$ & $\CEEight$ \\[0.5pt]
         \midrule
         $i$ & 2 & 2 & 1 & 1 & 2 & 2 & 1 & 1 \\[0.5pt]
         $j$ & 1 & 1 & 2 & 2 & 1 & 1 & 2 & 2 \\[0.5pt]
         $\mathrm{sgn}(\varphi_{1x})$ & $+$ & $-$ & + & + & + & $-$ & $-$ & $-$ \\[0.5pt]
         $\mathrm{sgn}(\varphi_{2x})$ & $+$ & $+$ & $-$ & + & $-$ & $-$ & $-$ & + \\[0.5pt]
         $\mathrm{sgn}(\mu)$ & $+$ & $-$ & $-$ & + & $-$ & $+$ & $+$ & $-$ \\[0.5pt]
         \bottomrule
    \end{tabular}
    \caption{Tabulated relationships between experimental circuit parameters---including their indices and signs---and the CE notation introduced in Sec. \ref{sec:ControlErasureProtocolSetup}. $+ \; (-)$ corresponds to a positive (negative) sign of a particular circuit parameter.}
    \label{tab:table of CEs and notation}
\end{table*}
\end{document}